\documentclass[twocolumn,showpacs,aps,prl,letterpaper]{revtex4}
\usepackage{graphicx}
\usepackage{dcolumn}
\usepackage{amsmath}
\usepackage{epsfig}

\RequirePackage{xspace}

\begin{document}

\title{\large \bfseries \boldmath Partial wave analysis of
$\psi'\to\gamma\chi_{c0}\rightarrow \gamma p K^- \overline{\Lambda}$
\\being used for searching for baryon resonance}
\author{Xian-Wei Kang$^{1,2}$} \email{kangxw@ihep.ac.cn}
\author{Hai-Bo Li$^2$}\email{lihb@ihep.ac.cn}
\author{Gong-Ru Lu$^1$}
\author{Bing-Song Zou$^2$}
\affiliation{$^1$Department of Physics, Henan Normal University,
Xinxiang 453007, China.\\$^2$Institute of High Energy Physics,
P.O.Box 918, Beijing  100049, China.}

\begin{abstract}
The abundant $\psi'$ events have been collected at the Beijing
Electron Positron Collider-II (BEPCII) that could undoubtedly
provide us with a great opportunity to study the more attractive
charmonium decays. As has been noticed before, in the process of
$J/\psi$ decaying to the baryonic final states, $p K^-
\overline{\Lambda}$, the evident $\Lambda^*$ and $N^*$ bands have
been observed. Similarly, by using the product of $\chi_{cJ}$ from
$\psi'$ radiative decay, we may confirm it or find some extra new
resonances. $\chi_{c0}$'s data samples will be more than
$\chi_{c1,2}$ taking into account the larger branching ratio of
$\psi'\to\gamma\chi_{c0}$. Here, we provide explicit partial wave
analysis formulae for the very interesting channel
$\psi'\to\gamma\chi_{c0}\rightarrow \gamma p K^-
\overline{\Lambda}$.
\end{abstract}


\maketitle


\section{Introduction}
In experiment at BESIII, about $10\times10^9 J/\Psi$ and
$3\times10^9 \psi'$ events can be collected per year's running
according to the designed luminosity of BEPCII in Beijing
\cite{besiii} \cite{bepcii}. These large data samples will provide
great opportunities to study the attractive $\chi_{cJ}$ decay and
some hyperon interactions. The product of $\chi_{cJ}$ in $\psi'$
radiative decays may provide useful information on two-gluon
hadronization dynamics and glueball decays. On the other side, the
radiative decays of $\psi'\to\gamma\chi_{cJ}$ are expected to be
dominated by electric dipole (E1) transitions, with higher
multipoles suppressed by powers of photon energy divided by quark
mass \cite{Karl} and searching for contributions of higher
multipoles is promissing. The possibility of anomalous magnetic
moments of higher quark being lager than those for light ones may
exist \cite{Geffen}. Thus,$\chi_{cJ}$ decays contain abundant
interesting physics.

In this paper, the idea is motivated by that there may exist some
baryon resonances in the process of $\chi_{cJ}$ decays to baryons.
In fact, in our preliminary Monte Carlo study, we found an explicit
band structure, most possibly $\Lambda(1520)$, so it necessitate
this PWA. In order to get more useful information about the
resonance properties such as $J^{PC}$ quantum numbers, mass, width,
production and decay rates,{\itshape{etc.}}, partial wave analysis
(PWA) are necessary. PWA is an effective method for analyzing the
experimental data of hadron spectrums. There are two methods of PWA,
one is based on the covariant tensor (also named Rarita-Schwinger)
formalism \cite{Schwinger} and the other is based on the original
helicity formalism \cite{Jackson} \cite{Jacob} and one covariant
helicity one developed by Chung \cite{Chung1} \cite{Chung2}. For a
more basic exposition, the reader may wish to consult the CERN
Yellow Report \cite{yellow report}. \,Ref.\cite{Filippini} showed
the connection between the covariant tensor formalism and helicity
one. In this short paper, we will pay more attention to the more
popular one: covariant tensor format also append the helicity one
for the specific process for $\Lambda(1520)$.

The organization of the paper is as follows. In Sec.I, the general
formalism is given. In Sec.II, we will present the covariant tensor
amplitude for $\psi'\to\gamma\chi_{c0}\rightarrow \gamma P K^-
\overline{\Lambda}$. In Sec.III,the corresponding helicity formula
are provides. At last, In Sec.IV,there is the conclusion.

\section{General formalism}
 \label{Sec:formalism}
In this part,the general formalism which will be used in the
following have been mentioned in Ref.\cite{psi decay to mesons}
\cite{N*NM couplings} \cite{gamma chicJ}, including formalism for
$\psi$ radiative decay to mesons(denoted by $M$),$M\to N^*
N$,$N^*\to N M$,where $N^*$ and $N$ has the half integer spin.

As discussed in Ref.\cite{psi decay to mesons},we denote the $\psi$
polarization four-vector by $\psi_{\mu}(m_1)$ and the polarization
vector of the photon by $e_{\nu}(m_2)$.Then the general form for the
decay  amplitude is
\begin{equation}
A=\psi_{\mu}(m_1)e^*_{\nu}(m_2)A^{\mu\nu}=\psi_{\mu}(m_1)e^*_{\nu}(m_2)\underset{i}{\sum}\Lambda_i
U^{\mu\nu}_i
\end{equation}
here, $U^{\mu}_{i}$ is the $i-$th partial wave amplitude with
coupling strength determined by a complex parameter $\Lambda_i$.
 Because of massless properties,there are two additional
conditions,$(1)$ the usual orthogonality condition
$e_{\nu}q^{\nu}=0$,where $q$ is the photon momentum;$(2)$ gauge
invariance condition (assuming the Coulomb gauge in $\psi$ rest
system) $e_{\nu}p^{\nu}_{\psi}=0$,where $P_{\psi}$ is the momentum
of vector meson $\psi$.Then we have
\begin{eqnarray}
\underset{m}{\sum}e^*_{\mu}(m)e_{\nu}(m)&=&-g_{\mu\nu}+\frac{q_{\mu}K_{\nu}+K_{\mu}q_{\nu}}{q\cdot
K}-\frac{K \cdot K}{(q \cdot K)^2}q_{\mu}q_{\nu}\nonumber\\ &\equiv&
-g_{\mu\nu}^{(\perp\perp)}
\end{eqnarray}
with $K=p_{\psi}-q $ and $ e_{\nu}K^{\nu}=0.$.To compute the
differential cross section,we need an expression for $|A|^2$ ,the
square modulus of the decay amplitude,which gives the decay
probability of a certain configuration should be independent of any
particular frame.Thus the radiative cross section is :
\begin{eqnarray}
\frac{d\sigma}{d\Phi_n}&=&\frac{1}{2}\sum^2_{m_1=1}\sum^2_{m_2=1}\psi_{\mu}(m_1)e^*_{\nu}(m_2)A^{\mu\nu}\psi^*_{\mu'}(m_1)e_{\nu'}(m_2)A^{*\mu'\nu'}
\nonumber\\
&=&-\frac{1}{2}\sum^2_{m_1=2}\psi_{\mu}(m_1)\psi_{\mu'}(m_1)g^{(\perp
\perp)}_{\nu\nu'}A^{\mu\nu}A^{*\mu'\nu'}\nonumber\\
&=&-\frac{1}{2}\sum^2_{\mu=1}A^{\mu\nu}g^{(\perp\perp)}_{\nu\nu'}A^{*\mu\nu'}\nonumber\\
&=&-\frac{1}{2}\sum_{i,j}\Lambda_i\Lambda_j^*\sum^2_{\mu=1}U^{\mu\nu}_ig^{(\perp\perp)}_{\nu\nu'}U_j^{*\mu\nu'}\equiv\sum_{i,j}P_{ij}\cdot
F_{ij}
\end{eqnarray}
with definition
\begin{eqnarray}
P_{ij}&=&P_{ji}^*=\Lambda_i\Lambda_j^*, \\F_{ij}&=&F_{ji}^*
=-\frac{1}{2}\sum^2_{\mu=1}=-\frac{1}{2}U^{\mu\nu}_i
g^{(\perp\perp)}_{\nu\nu'}U_j^{*\mu\nu'}.
\end{eqnarray}
note the relation
\begin{equation}
\sum^2_{m=1}\psi_{\mu}(m)\psi_{\mu'}^*(m)=\delta_{\mu\mu'}(\delta_{\mu1}+\delta_{\mu2}).
\end{equation}
The partial wave amplitude $U$ in the covariant Rarita-Schwinger
tensor formalism \cite{Schwinger} can be constructed by using pure
orbital angular momentum covariant tensor
$\widetilde{t}^{(L)}_{\mu_1\mu_2\cdots\mu_L}$ and covariant spin
wave functions $\phi_{\mu_1\mu_2\cdots\mu_S}$ together with the
metric tensor $g^{\mu\nu}$, the totally antisymmetric Levi-Civita
tensor $\epsilon_{\mu\nu\lambda\sigma}$ and the four momenta of
participating particles.For a process $a\to bc$ ,if there exists a
relative orbital angular momentum $L_{bc}$ between the particle $b$
and $c$ ,then the pure orbital angular momentum $L_{bc}$ state can
be represented by the covariant tensor wave function
$\widetilde{t}^{(L)}_{\mu_1\mu_2\cdots\mu_L}$  which is built of the
relative momentum.Here,we list the amplitude for pure $S-,P-,D-,$
and $F-$ wave orbital angular momentum:
\begin{eqnarray}
\widetilde{t}^{(0)}&=&1,\\
\widetilde{t}^{(1)}_{\mu}&=&\widetilde{g}_{\mu\nu}(p_a)r^{\nu}B_1(Q_{abc})\equiv
\widetilde{r}^{\mu}B_1(Q_{abc}),\\
 \widetilde{t}^{(2)}_{\mu\nu}&=&[\widetilde{r}^{\mu}\widetilde{r}^{\nu}-\frac{1}{3}(\widetilde{r}\cdot\widetilde{r})\widetilde{g}_{\mu\nu}(p_a))]B_2(Q_{abc}),
\end{eqnarray}
\begin{align}
 \widetilde{t}^{(3)}_{\mu\nu\lambda}&=[\widetilde{r}_{\mu}\widetilde{r}_{\nu}\widetilde{r}_{\lambda}-\frac{1}{5}(\widetilde{r}\cdot\widetilde{r})(\widetilde{g}_{\mu\nu}(p_a)\widetilde{r}_{\lambda}\nonumber \\
  &\quad+\widetilde{g}_{\nu\lambda}(p_a)\widetilde{r}_{\mu}+\widetilde{g}_{\lambda\mu}(p_a)\widetilde{r}_{\nu})]B_3(Q_{abc}),
\end{align}
where $r=p_b-p_c$ is the relative momentum of the two decay products
in the parent particle rest
frame,$\widetilde{r}\cdot\widetilde{r}=\vec{r}\cdot\vec{r}$ where
$\vec{r}$ is the magnitude of three-vector , with
\begin{equation}
\widetilde{g}_{\mu\nu}(p_a)=-g_{\mu\nu}+\frac{p_{a\mu}p_{a\nu}}{p_a^2}
\end{equation} which is the vector boson polarization sum relation,
and
\begin{equation}
Q^2_{abc}=\frac{(s_a+s_b-s_c)^2}{4s_a}-s_b
\end{equation} where $s_a=E_a^2-p_a^2$ and
 $B_l(Q_{abc})$ is the Blatt-Weisskopf barrier
factor \cite{Hippel},explicitly,
\begin{align}
B_1(Q_{abc})=&\sqrt{\frac{2}{Q_{abc}^2}+Q_0^2}\\
B_2(Q_{abc})=&\sqrt{\frac{13}{Q_{abc}^4}+3Q_{abc}^2Q_0^2+9Q_0^4},\\
B_3(Q_{abc})=&\sqrt{\frac{277}{Q_{abc}^6}+6Q_{abc}^4Q_0^2+45Q_{abc}^2Q_0^4+225Q_0^6}
\end{align}
Here $Q_0$ is a hadron scale parameter $Q_0=0.197321/R GeV/c$,in
which $R$ is the radius of the centrifugal barrier in fm.

  If $a$ is an intermediate resonance decaying into
  $bc$,one needs to introduce into the amplitude a Breit-Wigner
  propergator \cite{primer}
  \begin{equation}
  f^{(a)}_{(bc)}=\frac{1}{m_a^2-s_{bc}-im_a\Gamma_a}
  \end{equation}
In this equation,$s_{bc}=(p_b+p_c)^2$ is the invariant mass-squared
of $b$ and $c$; $m_a,\Gamma_a$ are the resonance mass and width.

  Additionally,some expressions depend also on the total momentum of
  the $ij$ pair,$p_{(ij)}=p_i+p_j$.When one wants to combine two
  angular momenta $j_b$ and $j_c$ into a total angular momentum
  $j_a$,if $j_a+j_b+j_c$ is an odd number,then a combination
  $\epsilon_{\mu\nu\lambda\sigma}p^{\mu}_{a}$ with $p_a$ the
  momentum of the parent particle is needed,otherwise it is not
  needed.

For a given hadronic decay process $A\rightarrow BC$ (B,C are
fermions),in the $L-S$ scheme on hadronic level,the initial state is
described by its $4-$ momentum $P_{\mu}$ and its spin state
$S_A$,the final state is described by the relative orbital angular
momentum state of $BC$ system and their spin state $(S_B,S_C)$.The
spin states $(S_A,S_B,S_C)$ can be well represented by the
relativistic Rarita-Schwinger spin wave functions for particles of
arbitrary spin.As is well known that,spin-$\frac{1}{2}$ wavefunction
is the standard Dirac spinor $U(p,s)$ and $V(p,s)$ ;spin-$1$
wavefunction is the standard spin-$1$ polarization four-vector
$\epsilon^{\mu}(p,s)$ for particle with momentum $p$ and spin
projection $s$.(1)For the case of A as a meson,B as $N^*$ with spin
$n+\frac{1}{2}$ and C as $\overline{N}$ with spin $\frac{1}{2}$ ,the
total spin of BC $(S_{BC})$ can be either $n$ or $n+1$. The two
$S_{BC}$ states can be represented as \cite{N*NM couplings}
\begin{align}
\psi^{(n)}_{\mu_1\mu_2\cdots\mu_n}=&\quad
\bar{u}_{\mu_1\mu2\cdots\mu_n}(p_B,s_B)\gamma_5
v(p_C,s_C),\\
\Psi^{(n+1)}_{\mu_1\mu_2\cdots\mu_{n+1}}=&\quad
\bar{u}_{\mu_1\mu2\cdots\mu_n}(\gamma_{\mu_{n+1}}-\frac{r_{\mu_{n+1}}}{m_A+m_B+m_C}v(p_c,s_C))
\nonumber\\&+(\mu_1\leftrightarrow\mu_{n+1})+\cdots+(\mu_{n}\leftrightarrow\mu_{n+1})
\end{align}
(2)For the case of A as $N^*$ with spin $n+\frac{1}{2}$,B as $N$ and
C as a meson,one needs to couple $-S_A$ and $S_B$ first to get
$S_{AB}=-S_A+S_B$ states,which are
\begin{align}
\phi^{(n)}_{\mu_1\mu_2\cdots\mu_n}=&\bar{u}(p_b,s_B)u_{\mu_1\mu_2\cdots\mu_n}(p_A,s_A),\\
\Phi^{n+1}_{\mu_1\mu_2\cdots\mu_n}=&\bar{u}(p_b,s_B)\gamma_5\widetilde{\gamma}_{\mu_1\mu_2\cdots\mu_n}(p_A,s_A)\nonumber\\
&+(\mu_1\leftrightarrow\mu_{n+1})+\cdots+(\mu_{n}\leftrightarrow\mu_{n+1})
\end{align}

Up to now,we have introduced all knowledges for constructing the
covariant tensor amplitude.In the concrete case,the P parity
conservation may be applied,which expression is
\begin{equation}\label{parity}
\eta_A=\eta_B\eta_C(-1)^{L}
\end{equation}
where $\eta_A$,$\eta_B$ and $\eta_C$ are the intrinsic parities of
particles A, B, and C, respectively.From this relation,L can be even
or odd for one case,which guarantee a pure L final state,which is
the soul of covariant $L-S$ coupling scheme.

\section{analysis for $\psi'\rightarrow \gamma \chi_{c0}\rightarrow \gamma P K^- \overline{\Lambda}$}
From now on,we denote $P$,$K^-$,$\overline{\Lambda}$ by number
$1,2,3$.Firstly,for $\psi'\to \gamma \chi_{c0}$,from the helicity
formalism,it is easy to show that there is only one independent
amplitude for $\psi'$ radiative deca
y to a spin $0$ meson.Hense,the
amplitude is
\begin{equation}
U^{\mu\nu}_{\gamma \chi_{c0}}=g^{\mu\nu}f^{(\chi_{c0})}.
\end{equation}
For sequential $\chi_{c0}$ decay,there may be the following modes:
$\chi_{c0}\to\Lambda_{x}\overline{\Lambda},\Lambda_x\rightarrow P
K^-$,where $\Lambda_{x}$ can be\\[0.2cm]
$\Lambda(1520)\frac{3}{2}^-,\Lambda(1600)\frac{1}{2}^+,
\Lambda(1670)\frac{1}{2}^-,\Lambda(1690)\frac{3}{2}^-,\Lambda(1800)\frac{1}{2}^-,\\
\Lambda(1810)\frac{1}{2}^+,\Lambda(1820)\frac{5}{2}^+,\Lambda(1830)\frac{5}{2}^-,\Lambda(1890)\frac{3}{2}^+,
\Lambda(2100)\frac{7}{2}^-,\\ \Lambda(2110)\frac{5}{2}^+;$\\[0.2cm]
$\chi_{c0}\rightarrow\overline{N}\overline{\Lambda},\overline{N}\rightarrow
\overline{\Lambda}K^-$,where $\overline{N}$ is the anti-partner of
hyperon $N$,with that $N$ can be\\[0.2cm]
$N(1650)\frac{1}{2}^-$,$N(1675)\frac{5}{2}^-$,$N(1700)\frac{3}{2}^-$,$N(1710)\frac{1}{2}^+$
or $N(1720)\frac{3}{2}^+$.\\[0.2cm] Another possibility that $P
\overline{\Lambda}$ may be generated from an unknown intermediate
resonance $K_x$ is also taken into account.Amplitudes for up to
$K_x's$ spin-4 are given. For $\Lambda_x$ being
$\Lambda(1520)\frac{3}{2}^-$,the total spin of $\Lambda(1520)$ and
$\overline{\Lambda}\frac{1}{2}^-$ can be $1$ or $2$,corresponding to
the $P$ wave and $D$ wave respectively,because of the special
property of spin-$0$ of $\chi_{c0}$.The parity relation
\eqref{parity} makes $P$ wave impossible.Considering that this
channel is recognized as a meson decaying to two fermions,now one
can write the covariant amplitude as
\begin{equation}
\Phi_{(\mu\nu)}^{(2)}\widetilde{t}^{(2)\mu\nu}.
\end{equation}
And then considering $\Lambda(1520)\rightarrow P K^-$,the total spin
of particle $1$ and $2$ can only be $\frac{1}{2}$,corresponding to
the $P$ wave and $D$ wave,after the parity formula being applied,$L$
must be $2$.As it belongs to a fermion decay to a fermion and a
meson,the covariant amplitude can be expressed as
$\Phi^{(2)_{\mu\nu}}\widetilde{t}^{(2)\mu\nu}$.

Here,we list all the amplitudes for the whole decay chain
$\psi'\to\gamma\chi_{c0},\chi_{c0}\rightarrow\Lambda_x\overline{\Lambda},\Lambda_x\rightarrow
P K^-$ up to spin-$\frac{7}{2}$ for $\Lambda_x$ according to the
above principles.
\begin{align}
\Lambda_x(\frac{1}{2}^+)\qquad
U^{\mu\nu}&=g^{\mu\nu}f^{(\chi_{c0})}_{(123)}\Psi^{(1)}_{\lambda}\widetilde{t}^{(1)\lambda}\Phi^{(1)}_{\sigma}\widetilde{t}^{(1)\sigma}
\\ \Lambda_x(\frac{1}{2}^-)\qquad U^{\mu\nu}&=g^{\mu\nu}f^{\chi_{c0}}_{(123)}\psi^{(0)}\phi^{(0)}
\\\Lambda_x(\frac{3}{2}^+)\qquad
U^{\mu\nu}&=g^{\mu\nu}f^{(\chi_{c0})}_{(123)}\psi^{(1)}_{\lambda}\widetilde{t}^{(1)\lambda}\phi^{(1)}_{\sigma}\widetilde{t}^{(1)\sigma}
\\\Lambda_x(\frac{3}{2}^-)\qquad U^{\mu\nu}&=g^{\mu\nu}f^{(\chi_{c0})}_{(123)}\Psi^{(2)}_{\lambda\delta}\widetilde{t}^{(2)\lambda\delta}\Phi^{(2)}_{\rho\sigma}\widetilde{t}^{(2)\rho\sigma}
\\\Lambda_x(\frac{5}{2}^+)\qquad U^{\mu\nu}&=g^{\mu\nu}f^{(\chi_{c0})}_{(123)}\Psi^{(3)}_{\lambda\delta\beta}\widetilde{t}^{(3)\lambda\delta\beta}\Phi^{(3)}_{\rho\sigma\eta}\widetilde{t}^{(3)\rho\sigma\eta}
\\\Lambda_x(\frac{5}{2}^-)\qquad U^{\mu\nu}&=g^{\mu\nu}f^{(\chi_{c0})}_{(123)}\psi^{(2)}_{\lambda\delta}\widetilde{t}^{(2)\lambda\delta}\phi^{(2)}_{\rho\sigma}\widetilde{t}^{(2)\rho\sigma}
\\\Lambda_x(\frac{7}{2}^+)\qquad U^{\mu\nu}&=g^{\mu\nu}f^{(\chi_{c0})}_{(123)}\psi^{(3)}_{\lambda\delta\beta}\widetilde{t}^{(3)\lambda\delta\beta}\phi^{(3)}_{\rho\sigma\eta}\widetilde{t}^{(3)\rho\sigma\eta}
\\\Lambda_x(\frac{7}{2}^-)\qquad U^{\mu\nu}&=g^{\mu\nu}f^{(\chi_{c0})}_{(123)}\Psi^{(4)}_{\lambda\delta\beta\xi}\widetilde{t}^{(4)\lambda\delta\beta\xi}\Phi^{(4)}_{\rho\sigma\eta\zeta}\widetilde{t}^{(4)\rho\sigma\eta\zeta}
\end{align}note that $\widetilde{t}^{(0)}=1$.For channel $\chi_{c0}\rightarrow
\overline{N_x}P,\overline{N_x}\rightarrow K^-\overline{\Lambda}$,we
can imitate the amplitude up to spin-$\frac{7}{2}$ for $N_x$ without
any difficulty,even through the highest spin for $N_x$ decaying into
$K^-\Lambda$ can only be $\frac{5}{2}$ presently \cite{pdg2008}.
\begin{align}
\overline{N_x}(\frac{1}{2}^+)\qquad
U^{\mu\nu}&=g^{\mu\nu}f^{(\chi_{c0})}_{(123)}\psi^{(0)}\phi^{(0)}
\\\overline{N_x}(\frac{1}{2}^-)\qquad U^{\mu\nu}&=g^{\mu\nu}f^{(\chi_{c0})}_{(123)}\Psi^{(1)}_{\lambda}\widetilde{t}^{(1)\lambda}\Phi^{(1)}_{\sigma}\widetilde{t}^{(1)\sigma}
\\\overline{N_x}(\frac{3}{2}^+)\qquad U^{\mu\nu}&=g^{\mu\nu}f^{(\chi_{c0})}_{(123)}\Psi^{(2)}_{\lambda\delta}\widetilde{t}^{(2)\lambda\delta}\Phi^{(2)}_{\rho\sigma}\widetilde{t}^{(2)\rho\sigma}
\\\overline{N_x}(\frac{3}{2}^-)\qquad
U^{\mu\nu}&=g^{\mu\nu}f^{(\chi_{c0})}_{(123)}\psi^{(1)}_{\lambda}\widetilde{t}^{(1)\lambda}\phi^{(1)}_{\sigma}\widetilde{t}^{(1)\sigma}
\\\overline{N_x}(\frac{5}{2}^+)\qquad U^{\mu\nu}&=g^{\mu\nu}f^{(\chi_{c0})}_{(123)}\psi^{(2)}_{\lambda\delta}\widetilde{t}^{(2)\lambda\delta}\phi^{(2)}_{\rho\sigma}\widetilde{t}^{(2)\rho\sigma}
\\\overline{N_x}(\frac{5}{2}^-)\qquad U^{\mu\nu}&=g^{\mu\nu}f^{(\chi_{c0})}_{(123)}\Psi^{(3)}_{\lambda\delta\beta}\widetilde{t}^{(3)\lambda\delta\beta}\Phi^{(3)}_{\rho\sigma\eta}\widetilde{t}^{(3)\rho\sigma\eta}
\\\overline{N_x}(\frac{7}{2}^+)\qquad U^{\mu\nu}&=g^{\mu\nu}f^{(\chi_{c0})}_{(123)}\Psi^{(4)}_{\lambda\delta\beta\xi}\widetilde{t}^{(4)\lambda\delta\beta\xi}\Phi^{(4)}_{\rho\sigma\eta\zeta}\widetilde{t}^{(4)\rho\sigma\eta\zeta}
\\\overline{N_x}(\frac{7}{2}^-)\qquad U^{\mu\nu}&=g^{\mu\nu}f^{(\chi_{c0})}_{(123)}\psi^{(3)}_{\lambda\delta\beta}\widetilde{t}^{(3)\lambda\delta\beta}\phi^{(3)}_{\rho\sigma\eta}\widetilde{t}^{(3)\rho\sigma\eta}
\end{align}
For channel $\chi_{c0}\rightarrow K^+_x K^-,K^+_x\rightarrow P
\overline{\Lambda}$,the amplitudes are also
given.$J^{P}=0^+,1^-,2^+,3^-,4^+ \cdots $ are forbidden by the
parity relation\cite{Parity}.The partial wave amplitude is denoted
by $U^{\mu\nu}_{(LS)}$, $L,S$ means the orbital angular momentum
number and spin angular momentum number between $P$ and
$\overline{\Lambda}$.
\begin{align}
K_x^+(0^-)\qquad
U^{\mu\nu}&=g^{\mu\nu}f^{(\chi_{c0})}_{(123)}\psi^{(0)}
\\K_x^+(1^+)\qquad
U^{\mu\nu}_{(10)}&=g^{\mu\nu}f^{(\chi_{c0})}_{(123)}\psi^{(0)}\widetilde{T}^{(1)\sigma}\phi_{\sigma}\epsilon_{\lambda}\widetilde{t}^{(1)\lambda}
\\U^{\mu\nu}_{(11)}&=g^{\mu\nu}f^{(\chi_{c0})}_{(123)}\epsilon^{\rho\sigma\eta\zeta}p_{K_x\rho}\epsilon_{\sigma}\widetilde{t}^{(1)}_{\eta}\Psi^{(1)}_{\zeta}\widetilde{T}^{(1)\sigma}\phi_{\sigma}
\\K_x^+(2^-)\qquad
U^{\mu\nu}_{(20)}&=g^{\mu\nu}f^{(\chi_{c0})}_{(123)}\widetilde{T}^{(2)\eta\zeta}\phi_{\eta\zeta}\widetilde{t}^{(2)\rho\sigma}\psi^{(0)}\phi_{\rho\sigma}\nonumber\\
&=g^{\mu\nu}f^{(\chi_{c0})}_{(123)}P^{(2)}_{\rho\sigma\eta\zeta}\psi^{(0)}\widetilde{T}^{(2)\eta\zeta}\widetilde{t}^{(2)\rho\sigma}
\\U^{\mu\nu}_{(21)}&=g^{\mu\nu}f^{(\chi_{c0})}_{(123)}\epsilon^{\rho\sigma\eta\zeta}p_{K_x\rho}\widetilde{T}^{(2)\beta\lambda}\phi_{\beta\lambda}\widetilde{t}^{(2)\iota}_{\sigma}\phi_{\iota\eta}\Psi^{(1)}_{\zeta}
\nonumber\\&=g^{\mu\nu}f^{(\chi_{c0})}_{(123)}P^{(2)}_{\beta\lambda\iota\eta}\epsilon^{\rho\sigma\eta\zeta}p_{K_x\rho}\widetilde{T}^{(2)\beta\lambda}\widetilde{t}^{(2)\iota}_{\sigma}\Psi^{(1)}_{\zeta}
\\K_x^+(3^+)\qquad
U^{\mu\nu}_{(30)}&=g^{\mu\nu}f^{(\chi_{c0})}_{(123)}\widetilde{T}^{(3)\lambda\delta\beta}\phi_{\lambda\delta\beta}\psi^{(0)}\widetilde{t}^{(3)\rho\sigma\eta}\phi_{\rho\sigma\eta}\nonumber
\\&=g^{\mu\nu}f^{(\chi_{c0})}_{(123)}P^{(3)}_{\lambda\delta\beta\rho\sigma\eta}\widetilde{T}^{(3)\lambda\delta\beta}\psi^{(0)}\widetilde{t}^{(3)\rho\sigma\eta}
\\U^{\mu\nu}_{(31)}&=g^{\mu\nu}f^{(\chi_{c0})}_{(123)}\widetilde{T}^{(3)\lambda\delta\beta}\phi_{\lambda\delta\beta}\epsilon^{\rho\sigma\eta\zeta}p_{K_x\rho}\widetilde{t}^{(3)\kappa\xi}_{\sigma}\Psi^{(1)}_{\eta}\cdot\nonumber
\\ &\quad\phi^{(3)}_{\zeta\kappa\xi}\nonumber\\
&=g^{\mu\nu}f^{(\chi_{c0})}_{(123)}P^{(3)}_{\lambda\delta\beta\zeta\kappa\xi}\widetilde{T}^{(3)\lambda\delta\beta}\epsilon^{\rho\sigma\eta\zeta}p_{K_x\rho}\widetilde{t}^{(3)\kappa\xi}_{\sigma}\Psi^{(1)}_{\eta}
\end{align}
In the above formulas,$\phi$ implies the spin wave functions for
$K_x$,for example,$\phi_{\sigma}$ corresponding to the
spin-1,$\phi_{\rho\sigma}$ corresponding to spin-2.There is a
general wave fuction for a particle of spin $J$ \cite{Chung1},which
is a rank-$J$ tensor
\begin{align}
\phi^{\alpha_1\alpha_2\cdots\alpha_J}(m)&=\underset{m_1m_2\cdots}{\sum}\langle
1m_11m_2|2n_1\rangle\langle
2n_11m_3|3n_2\rangle\nonumber\\
&\quad\cdots\langle
J-1n_{J-2}1m_J|Jm\rangle\phi^{\alpha_1}(m_1)\nonumber\\
&\quad\phi^{\alpha_2}(m_2)\cdots\phi^{\alpha_J}(m_J)
\end{align}
where $\phi^{\alpha}(m)$ is the familiar polarization four-vector of
spin-1 particle,
\begin{equation}
\phi^{\alpha}(1,-1)=\mp\frac{1}{\sqrt{2}}(0;1,\pm i,0),
\phi^{\alpha}(0)=(0;0,0,1).
\end{equation}
note the following useful relationship:
\begin{equation}
\phi(-m)=(-)^m\phi^*(m).
\end{equation}
It is best to illustrate these formulas with some examples.For
$J=1$,one finds that it reduces to identities for $\phi(1)$ and
$\phi(0)$.For $J=2$,one has
\begin{multline}
\qquad\qquad\phi^{\alpha\beta}(+2)=\phi^{\alpha}(1)\phi^{\beta}(1) \\
\phi^{\alpha\beta}(+1)=\frac{1}{\sqrt{2}}[\phi^{\alpha}(1)\phi^{\beta}(0)+\phi^{\alpha}(0)\phi^{\beta}(1)]\\
\phi^{\alpha\beta}(0)=\frac{1}{\sqrt{6}}[\phi^{\alpha}(1)\phi^{\beta}(-1)+\phi^{\alpha}(-1)\phi^{\beta}(1)+\sqrt{\frac{2}{3}}\phi^{\alpha}(0)\phi^{\beta}(0)]
\end{multline}
$\phi\cdot\phi$ is in reality its spin projection operator
$P^{(S)}$.
\begin{align}
P^{(2)}_{\rho\sigma\eta\zeta}(p_{K_x})&=\underset{m}{\sum}\phi_{\rho\sigma}(p_{K_x},m)\phi^*_{\eta\zeta}(p_{K_x,m})\nonumber\\&=\frac{1}{2}(\widetilde{g}_{\rho\eta}
\widetilde{g}_{\sigma\zeta}+\widetilde{g}_{\rho\zeta}\widetilde{g}_{\sigma\eta})-\frac{1}{3}\widetilde{g}_{\rho\sigma}\widetilde{g}_{\eta\zeta}
\\\quad P^{(3)}_{\lambda\delta\beta\zeta\kappa\xi}(p_{K_x})&=\underset{m}{\sum}\phi_{\lambda\delta\beta}(p_{K_x},m)\phi^*{\zeta\kappa\xi}(p_{K_x},m)\nonumber\\&=\frac{1}{6}(\widetilde{g}_{\lambda\zeta}\widetilde{g}_{\delta\kappa}\widetilde{g}_{\beta\xi}
+\widetilde{g}_{\lambda\zeta}\widetilde{g}_{\delta\xi}\widetilde{g}_{\beta\kappa}+\widetilde{g}_{\lambda\kappa}\widetilde{g}_{\delta\zeta}\widetilde{g}_{\beta\xi}
\nonumber\\&\quad\quad+\widetilde{g}_{\lambda\zeta}\widetilde{g}_{\delta\xi}\widetilde{g}_
{\beta\kappa}+\widetilde{g}_{\delta\xi}\widetilde{g}_{\beta\zeta}\widetilde{g}_{\lambda\xi}+\widetilde{g}_{\delta\kappa}\widetilde{g}_{\lambda\xi}\widetilde{g}_{\beta\zeta})
\nonumber\\&\quad-\frac{1}{15}(\widetilde{g}_{\lambda\delta}\widetilde{g}_{\zeta\kappa}\widetilde{g}_{\beta\xi}+\widetilde{g}_{\lambda\delta}\widetilde{g}_{\kappa\xi}\widetilde{g}_{\beta\zeta}+\widetilde{g}_{\lambda\delta}\widetilde{g}_{\zeta\xi}\widetilde{g}_{\beta\kappa}
\nonumber\\&\quad\quad+\widetilde{g}_{\lambda\beta}\widetilde{g}_{\zeta\xi}\widetilde{g}_{\delta\kappa}+\widetilde{g}_{\lambda\beta}\widetilde{g}_{\zeta\kappa}\widetilde{g}_{\delta\xi}
+\widetilde{g}_{\lambda\beta}\widetilde{g}_{\kappa\xi}\widetilde{g}_{\delta\zeta}
\nonumber\\&\quad\quad+\widetilde{g}_{\delta\beta}\widetilde{g}_{\kappa\xi}\widetilde{g}_{\lambda\kappa}+\widetilde{g}_{\delta\beta}\widetilde{g}_{\zeta\kappa}\widetilde{g}_{\lambda\xi}+\widetilde{g}_{\delta\beta}\widetilde{g}_{\kappa\xi}\widetilde{g}_{\lambda\kappa})
\end{align}
So far,we have given the covariant tensor amplitude formula for the
process $\psi'\to\gamma\chi_{c0}\rightarrow \gamma P K^-
\overline{\Lambda}$.
\section{helicity formula}
For completeness,we also give the helicity format in comparison with
tensor formula.Helicity formalism has an explicit advantage, the
angular dependence can be easily seen.In this section ,we will give
the amplitude for
$\chi_{c0}\to\Lambda(1520)\overline{\Lambda},\Lambda(1520)\to P
K^-,\overline{\Lambda}\to\overline{P}\pi^+$.$\Lambda(1520)$ the most
possible resononse,has been mentioned above. Firstly,we want to
introduce the general helicity formula expression.Consider a state
with spin(parity) $=J(\eta_J)$ decaying into two states with
$S(\eta_s)$ and $\sigma(\eta_\sigma)$.The decay amplitudes are
given,in the rest frame of $J$ \cite{helicity guide} \cite{Jacob},
\begin{equation}
\mathcal{M}^{J\to
s\sigma}_{\lambda\nu}=\sqrt{\frac{2J+1}{4\pi}}D^{J*}_{M\delta}(\phi,\theta,0)H^J_{\lambda\nu},
\end{equation}
where $\lambda$ and $\nu$ are the helicities of the two final state
particles $s$ and $\sigma$ with $\delta=\lambda-\nu$.The symbol $M$
stands for the $z$ component of the spin $J$ in a coordinate system
fixed by production process.The helicities $\lambda$ and $\nu$ are
rotational invariants by definition.The direction of the break-up
momentum of the decaying particle $s$ is given by the angles
$\theta$ and $\phi$ in the $J$ rest frame.Let $\hat{x},\hat{y}$ and
$\hat{z}$ be the coordinate system fixed in the $J$ rest frame.It is
important to recognize,for applications to sequential decays,the
exact nature of the body-fixed (helicity) coordinate system implied
by the arguments of the $D$ function given above.Let
$\hat{x}_h,\hat{y}_h$ and $\hat{z}_h$ be the helicity coordinate
system fixed by the $s$ decay.Then by definition $\hat{z}_h$
describes the direction of $s$ in the $J$ rest frame (also termed
the helicity axis) and the $y$ axis is given by
$\hat{y}_h=\hat{z}\times\hat{z}_h$ and
$\hat{x}_h=\hat{y}_h\times\hat{z}_h$.Parity conservation in the
decay leads to the relationship
\begin{equation}
H^{J}_{\lambda\nu}=\eta_J\eta_s\eta_\sigma(-)^{J-s-\sigma}H^{J}_{-\lambda-\nu}
\end{equation}
Let us consider a full process $A\to B+C$ where $B$ and $C$ are also
unstable particles decaying to $B_1+B_2$ and $C_1+C_2$
respectively.The decay amplitude is simply \cite{PhD thesis}
\begin{align}
\mathcal{M}(\lambda_{B_1},\lambda_{B_2},\lambda_{C_1},\lambda_{C_2})=\underset{\lambda_B,\lambda_C}{\sum}&
\mathcal{M}^{A\to B+C}_{\lambda_B,\lambda_C}\cdot \mathcal{M}^{B\to
B_1+B_2}_{\lambda_{B_1},\lambda_{B_2}}\cdot\nonumber
\\&\mathcal{M}^{C\to C_1+C_2}_{\lambda_{C_1},\lambda_{C_2}}
\end{align}
with
\begin{subequations}
\begin{align}
\mathcal{M}^{A\to
B+C}_{\lambda_B,\lambda_C}&=\sqrt{\frac{2J_A+1}{4\pi}}D^{J_A*}_{M_A,\lambda_B-\lambda_C}(\phi_A,\theta_A,0)H^A_{\lambda_B,\lambda_C},
\\ \mathcal{M}^{B\to
B_1+B_2}_{\lambda_{B_1},\lambda_{B_2}}&=\sqrt{\frac{2J_B+1}{4\pi}}D^{J_B*}_{\lambda_B,\lambda_{B_1}-\lambda_{B_2}}(\phi_B,\theta_B,0)H^B_{\lambda_{B_1},\lambda_{B_2}},
\\ \mathcal{M}^{C\to
C_1+C_2}_{\lambda_{C_1},\lambda_{C_2}}&=\sqrt{\frac{2J_C+1}{4\pi}}D^{J_C*}_{-\lambda_C,\lambda_{C_1}-\lambda_{C_2}}(\phi_C,\theta_C,0)H^C_{\lambda_{C_1},\lambda_{C_2}},
\label{sub3}
\end{align}
\end{subequations}
Please note in \eqref{sub3} the first subscript of $D^{J_C*}$ is
$-\lambda_C$ and NOT $\Lambda_C$ although it also gives the correct
result,because the quantization axis is along the direction of the
momentum of particle $B$ so that the spin-quantization projection
$M_C$ in the particle $C$ rest frame verifies $M_C=-\lambda_C$.

The unpolarized angular distribution is then given by averaging over
initial spins and by summing over final spins:
\begin{align}
\frac{d^3\Gamma}{\mathcal{N}d\Omega_A d\omega_B
d\Omega_C}=\frac{1}{2S_A+1}&\underset{\lambda_{B_1},\lambda_{C_1},\lambda_{B_2},\lambda_{C_2}}{\sum}\nonumber
\\&|\mathcal{M}(\lambda_{B_1},\lambda_{B_2},\lambda_{C_1},\lambda_{C_2})|^2.
\end{align}
where $\mathcal{N}$ is the normalization factor. Using the parity
conservation formula,one has
$H^{\overline{\Lambda}}_{\frac{1}{2}0}=H^{\overline{\Lambda}}_{-\frac{1}{2}0}$
and
$H^{\Lambda(1520)}_{\frac{1}{2}0}=-H^{\Lambda(1520)}_{-\frac{1}{2}0}$.By
applying the above amplitude expression,after a lengthy,tedious but
trivial evaluating,one can get the helicity amplitude,
\begin{align}\label{helicity distribution}
\frac{d^3\Gamma}{\mathcal{N}d\Omega_A d\omega_B
d\Omega_C}=&[\frac{3}{2}\cos^2\theta_{\Lambda(1520)}-\frac{3}{2}\cos\theta_{\Lambda(1520)}\nonumber
\\&+\frac{9}{2}\cos^2\theta_{\Lambda(1520)}\sin\theta_{\Lambda(1520)}\cos\phi_{\Lambda(1520)}\nonumber
\\&+\frac{\sqrt{3}}{2}\cos^2\theta_{\Lambda(1520)}\cos2\phi_{\Lambda(1520)}\nonumber
\\&-\frac{\sqrt{3}}{4}\cos\theta_{\Lambda(1520)}\cos2\phi_{\Lambda(1520)}\nonumber
\\&-\frac{3\sqrt{3}}{4}\cos2\phi_{\Lambda(1520)}+1]|H^{\overline{\Lambda}}_{\frac{1}{2}0}|^2H^{\Lambda(1520)}_{\frac{1}{2}0}|^2
\end{align}
where the subscript $\Lambda(1520)$ denotes that the angle defined
in the rest frame of $\Lambda(1520)$. After integrating
$\phi_{\Lambda(1520)}$'s from $[0,2\pi]$,the Eq.\eqref{helicity
distribution} becomes
\begin{equation}
\frac{d^3\Gamma}{\mathcal{N'}d\Omega_A d\omega_B
d\Omega_C}=\frac{3}{2}\cos^2\theta_{\Lambda(1520)}-\frac{3}{2}\cos\theta_{\Lambda(1520)}+1
\end{equation}
where
$\mathcal{N'}=\mathcal{N}|H^{\overline{\Lambda}}_{\frac{1}{2}0}|^2H^{\Lambda(1520)}_{\frac{1}{2}0}|^2$
is redefined normalization factor.
\section{conclusion}
In this short note,firstly,the relevant general tensor formalism hac
been introduced,and then give the covariant tensor amplitudes.At
last,for completeness and some experimental reasons,the helicity
amplitude expression has also been provided and a figure attached.
\section{acknowledgments}
The author acknowledges greatly helpful discussions with B.~S.
Zou.This work is supported in part by the National Natural Science
Foundation of China under contracts Nos. 10521003,10821063, the 100
Talents program of CAS, and the Knowledge Innovation Project of CAS
under contract Nos. U-612 and U-530 (IHEP).

\begin{figure*}\label{Fig}
\centering
\includegraphics[height=8cm,width=8cm]{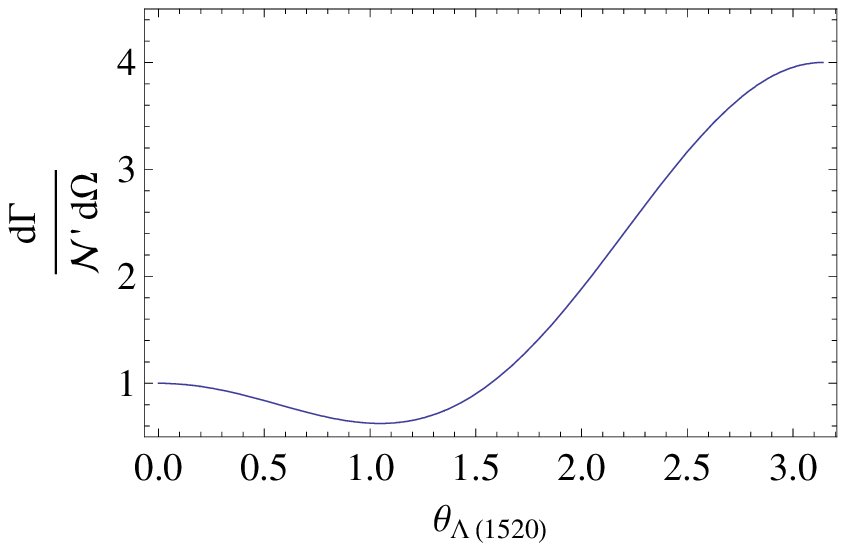}
\caption{The illustrative plot for the angular distribution of
$\chi_{c0}\to \Lambda(1520)\overline{\Lambda},\overline{\Lambda}\to
\overline{P}\pi^+,\Lambda(1520)\to P K^-$ in helicity format}
\end{figure*}

\end{document}